\documentstyle[12pt]{article}

\newcommand{\be}{\begin{equation}}
\newcommand{\ee}{\end{equation}}
\newtheorem{definition}{Definition }
\newtheorem{theorem}{Theorem }

\begin{document}
\begin{titlepage}
\title
{Integrability in the Hamiltonian Chern-Simons theory}
\author{A.Yu.Alekseev
\thanks{On leave of absence from Steklov Mathematical Institute,
Fontanka 27, St.Petersburg, Russia}
\thanks{Supported by Swedish Natural Science Research Council
(NFR) under the contract F-FU 06821-304}
\\\\
Institute of Theoretical Physics, Uppsala University,
\\ Box 803 S-75108, Uppsala, Sweden.\\\\
October 1993\\}
\date{\it Dedicated to L.D. Faddeev \\on the occasion of his $60^{\rm
th}$ birthday}
\maketitle \thispagestyle{empty}
\begin{abstract}
We consider the moduli space of flat connections
on the Riemann surface with marked points. The new efficient
parametrization is suggested
and used to construct an integrable model on the moduli space.
A family of commuting Hamiltonians is extracted from the trace of the
transfer matrix built from the Wilson line observables of the
Chern-Simons theory.  Our model appears to be gauge equivalent to XXZ
magnetic chain with finite number of sites.
\end{abstract}
\end{titlepage}

\section{Introduction}
\setcounter{equation}{0}
Quantization of the Chern-Simons theory in 3 dimensions became
the object of intensive investigations when the relation between the
Chern-Simons functional integral and the theory of knots had been
discovered \cite{Wt1}. Let us start with a brief description of the
model.

\subsection{Chern-Simons theory}
The Chern-Simons model corresponding
to the Lie algebra ${\cal G}$ is defined as a theory
of the gauge connection $A$ with the action

\begin{equation}
CS(A)=\frac{k}{4\pi}Tr\int_{\cal M}(AdA+\frac{2}{3}A^{3}), \label{CS}
\end{equation}
where the integration region $M$ is a 3-dimensional manifold. The
theory
(\ref{CS}) appears to be topological because the action is written in
terms
of differential forms and hence there is no metric dependence from
the
very beginning. In principle, metric may influence the theory through
regularization procedure but we don't touch these subtleties here.

The model (\ref{CS}) enjoys two symmetries: the gauge symmetry
(for integer values of $k$) and reparametrization symmetry. So, we
should construct observables in such a way that they
have the same symmetries as the action.
The simplest example is provided by the Wilson line observables

\begin{equation}
W_{I}(\Gamma)=tr_{I}Pexp\int_{\Gamma}A^{I}.
\end{equation}
Here $\Gamma$ is a closed curve in the manifold $M$ and the
connection
$A^{I}=A^{a}_{i}T^{a}_{I}dx^{i}$ is evaluated in the representation
$I$ of the
algebra ${\cal G}$. So in general we deal with evaluation of the
Wilson lines
correlator of the following type:

\begin{equation}
Z_{k}(I_{1},\dots,I_{n}) = \int DA e^{iCS(A)} W_{I_{1}}\dots
W_{I_{n}}.  \label{Wilpart}
\end{equation}

Let us simplify the problem and assume that at least locally the
manifold $M$ looks like
a cylinder $\Sigma\times R$, where $\Sigma$ is a Riemann surface.
Then
the problem may be treated in the framework of Hamiltonian mechanics.
We shall refer to the Chern-Simons model on the cylinder as to
Hamiltonian Chern-Simons model.

It is well-known that in the Hamiltonian CS theory the gauge field
$A$
is constrained by the flatness condition on the equal-time surface
$\Sigma$:

\begin{equation}
F = dA + A^{2} = 0.  \label{F=0}
\end{equation}
The reduced phase space may be obtained as a quotient of the space of
flat connections over the gauge group action

\begin{equation}
A^{g} = g A g^{-1} - dg g^{-1}.  \label{Ggtr}
\end{equation}
In this way the moduli space of flat connection appears.
If we have some Wilson lines intersecting $\Sigma$, the curvature may
develop $\delta$-function singularities at the intersection points.
In
this case we deal with the
moduli space of flat connections with marked points. Marked points
are
exactly those
where the Wilson lines intersect $\Sigma$.  Each marked point is
equipped with the representation of ${\cal G}$ carried by the
corresponding Wilson line.

In this paper I consider quantization of the Hamiltonian Chern-Simons
theory.
Let me briefly describe the plan of the paper.
Section 2 is devoted to the algebra of observables which may be
interpreted as quantized algebra of functions on the moduli space of
flat connections (moduli algebra). In this part I extensively use the
material of \cite{FoRo},\cite{AGS}.  In Section 3 I introduce a new
integrable model on the moduli space of flat connections. Being
topological the Chern-Simons theory itself has the Hamiltonian equal
to zero. However, one may look for the complete set of commuting
observables.A certain number of them may be constructed starting from
Wilson line observables. At this point the technique of Inverse
Scattering method \cite{F} appears to be useful.  I postpone to the
next paper more serious discussion of completeness of the commutative
family .  The equivalence of the integrable model of Section 3 and
and
XXZ magnetic chain is established in Section 4.  In Section 5 I
return
to the moduli algebra and construct its irreducible representations
for generic value of the deformation parameter $q$ (noninteger
coupling constant $k$ in the CS theory).

\subsection{Background}
Some basic references are collected in this subsection.

The main technical tool of this paper is the lattice simulation of
the
Hamiltonian
Chern-Simons theory.  The approach I am following here originates
from
the construction
of the lattice current algebra \cite{AFS} where the lattice
simulation
of the Wess-Zumino-Novikov-Witten (WZNW) model has been suggested.
The
current algebra may
be simulated on the finite lattice by the quadratic $R$-matrix
algebra. In this approach the relation to the Quantum groups is
especially transparent. Later the spectral dependent $L$-operator has
been introduced and the family of commuting integrals of motion
in the lattice WZNW model has been constructed \cite{FV} .  Here we
fulfil the same program
but for the different algebra.

The next important step towards the correct lattice approximation of
the CS theory has been made in \cite{FoRo} where the proper
discretization of
the moduli space of flat connections has been suggested. The idea is
to draw a graph
on the Riemann surface and consider the set of parallel transports
along the graph links
(link variables) instead of the two-dimensional gauge field.  It is
possible to introduce the
quadratic $R$-matrix algebra for the link variables consistent with
the standard quantization of the Chern-Simons model. In the simplest
case when the graph is just a polygon we get the same $R$-matrix
algebra as we had for the lattice version of the WZNW model. Because
of the topological nature of the CS theory its lattice simulation has
an
advantage in comparison with the lattice WZNW model.  The lattice CS
model is expected to give exactly the same results as the continuous
theory. In this connection it is worth mentioning that the similar
lattice simulation has been successfully used in the construction of
invariants of 3-dimensional manifolds \cite{TV}.

The $R$-matrix algebra of link variables in the lattice CS theory
belongs to the class of nonultralocal quadratic algebras. It means
that the variables assigned to different links do not necessarily
commute.  The general theory of nonultralocal quadratic algebras has
been developed in \cite{MF},\cite{Sem}.

Let me finish with the remark that the attempt to represent the CS
theory as a lattice gauge model has been described in \cite{Bou}.
The
link algebra in this model is ultralocal. As
a consequence the model does not enjoy gauge invariance. However, the
paper \cite{Bou}
includes many useful observations ({\em e.g.} the canonical integral
on the link algebra).

\subsection{Notations}
Here I introduce some useful notations.
First of all we refer to the finite dimensional algebra ${\cal G}$
and
its set of irreducible representations $\Im$. Particular irreducible
representations will
be usually denoted by $I_{1},\dots ,I_{n}\in \Im $.

We shall often deal with parallel transports defined by the flat
connection $A$ on the
Riemann surface. For a given closed curve $\Gamma$ we introduce the
monodromy

\begin{equation}
M_{\Gamma}=P exp\int_{\Gamma}A. \label{Mon}
\end{equation}
It is convenient to have the matrix $M$ in any representation $I\in
\Im $:

\begin{equation}
M^{I}_{\Gamma}=P exp\int_{\Gamma}A^{I}.  \label{MonI}
\end{equation}

The next important object which will be used to describe the moduli
algebra
is the quantum $R$-matrix. One can treat it as an element of the
tensor square
of the quantized universal enveloping algebra $U_{q}({\cal
G})^{\otimes 2}$.
If we evaluate the first multiplier of $R$ in the representation $I$
and the second
multiplier in the representation $J$, we get the numerical matrix
$R^{IJ}$. The set
of numerical $R$-matrices satisfies the quantum Yang-Baxter equation:

\begin{equation}
R_{12}^{IJ} R_{13}^{IK} R_{23}^{JK}=
R_{23}^{JK} R_{13}^{IK} R_{12}^{IJ}.  \label{qyb}
\end{equation}
The $R$-matrix corresponding to $U_{q}({\cal G})$ depends on the
deformation parameter $q$. For example, let us write the $4\times 4$
$R$-matrix corresponding to the algebra $sl(2)$:

\begin{equation}
R=\left (\matrix{
q & 0 & 0 & 0 \cr
0 & 1 & (q-q^{-1}) & 0 \cr
0 & 0 & 1 & 0 \cr
0 & 0 & 0 & q \cr
}\right ).  \label{44R}
\end{equation}
If we have a solution of (\ref{qyb}) it is always possible to get 3
more solutions in the following simple way. Let us redenote the
matrix
(\ref{44R}) by $R_{+}$ and consider $R_{+}^{-1}$,
$R_{-}$ and $R_{-}^{-1}$, where

\begin{equation}
R_{-}=P R_{+}^{-1} P.  \label{PRP}
\end{equation}
All of them are solutions of (\ref{qyb}). Here $P$ is the permutation
matrix
which exchanges two representation spaces in the tensor square.  The
$R$-matrices $R_{\pm}$ and $R_{\pm}^{-1}$ play the role of structure
constants in the $R$-matrix algebras.

It is worth mentioning that when the deformation parameter $q$ is
closed to unity, all $R$-matrices approach the unit matrix:

\begin{equation}
R_{\pm}=I+(q-1)r_{\pm}+\dots.  \label{Rr}
\end{equation}
Now we can turn to the description of the moduli algebra.

\section{Moduli algebra}
\setcounter{equation}{0}

One way to deal with the moduli algebra is to choose
a particular graph on the Riemann surface. There is no canonical
choice and we do it trying to get the the most economic description.
Let us choose a graph to be a bunch of circles $B$ intersecting at
the
only point
$P$. In this bunch we have two circles for each handle (corresponding
to
$a$- and $b$- cycles) and one circle for each marked point. We shall
denote the circles corresponding to the $i$'s handle by $a_{i}$ and
$b_{i}$ ($i = 1, \dots ,g$) and we shall use
symbols $m_{i}$ ($i = 1,\dots ,n$) for the circles surrounding marked
points.
We shall assume that the circles on $\Sigma$
are chosen in such a way that the only defining relation
in $\pi_{1}(\Sigma_{g,n})$ looks as

\begin{equation}
m_{1}\ldots m_{n} (a_{1}b_{1}^{-1}a_{1}^{-1}b_{1})\ldots
(a_{g}b_{g}^{-1}a_{g}^{-1}b_{g}) = id.
\label{abc}
\end{equation}
In particular, equation (\ref{abc}) fixes the cyclic order of links
incident to
the graph vertex $P$.

Now we can define the algebra $\cal F$ of link variables
corresponding
to the graph $B$. To each circle we assign the corresponding
monodromy
matrix.
Let us denote these matrices by $A_{i}$,$B_{i}$ and $M_{i}$ for $a$-,
$b$- and
$m$-circles. The set of monodromy matrices provides the
representation
of the
fundamental group $\pi_{1}(\Sigma_{g,n})$. It implies the relation

\begin{equation}
M_{1}\ldots M_{n} (A_{1}B_{1}^{-1}A_{1}^{-1}B_{1})\ldots
(A_{g}B_{g}^{-1}A_{g}^{-1}B_{g})= I
\label{ABC}
\end{equation}
imposed on the values of $A_{i}$,$B_{i}$ and $M_{i}$. In principle,
we
may introduce
a set of matrices for each link so that any monodromy (for example
$A_{1}$)
appears in any representation $I$ of the finite dimensional algebra
$\cal G$. We shall
denote corresponding matrices by $A^{I}_{i}$,$B^{I}_{i}$ and
$M^{I}_{i}$. Apparently, relation
(\ref{ABC}) holds true for any representation $I$.

We define the algebra $\cal F$ by the set of quadratic $R$-matrix
relations
for the matrix elements of monodromies. It is useful to introduce
notation $X_{i}$
referring to arbitrary monodromy matrix . So the symbol $X_{i}$ may
denote $A_{i}$,$B_{i}$
or $M_{i}$. It is convenient to introduce the partial order in the
set
of circles.  If the letter
$x_{i}$ appears from the left hand side of the letter $x_{j}$ in the
word (\ref{abc}), we can
express it as $i<j$. This definition should not be applied to $a$-
and
$b$- cycles
winding around the same handle as it gets ambiguous.

\begin{definition}
The algebra ${\cal F}_{g,n}$ is generated by the matrix elements of
the monodromy matrices
$A^{I}_{i}$,$B^{I}_{i}$ and $M^{I}_{i}$ subject to the following set
of quadratic relations:

\begin{equation}
R_{-}X_{i}^{1}R_{-}^{-1}X_{i}^{2}=X_{i}^{2}R_{+}X_{i}^{1}R_{+}^{-1}
\label{X=X}
\end{equation}
for any monodromy matrix $X_{i}$;

\begin{equation}
R_{+}X_{i}^{1}R_{+}^{-1}X_{j}^{2}=X_{j}^{2}R_{+}X_{i}^{1}R_{+}^{-1}
\label{X>X}
\end{equation}
for the matrix elements of two different monodromies $X_{i}$ and
$X_{j}$ if $i<j$;

\begin{equation}
R_{+}A_{i}^{1}R_{-}^{-1}B_{i}^{2}=B_{i}^{2}R_{+}A_{i}^{1}R_{+}^{-1}
\label{AB}
\end{equation}
for monodromies $A_{i}$ and $B_{i}$ corresponding to $a$- and
$b$-cycles
of the same handle.
\end{definition}

Let us make several comments concerning this definition.

{\em Remark 1.} The classical analogue of the algebra ${\cal F}$ may
be easily
defined if we consider the limit when the deformation parameter $q$
which enters
all $R$-matrices tends to 1.  Exchange relations
(\ref{X=X}--\ref{AB})
will be
replaced by quadratic $r$-matrix Poisson brackets:

\begin{equation}
\{X_{i}^{1}, X_{i}^{2}\} = -r_{-}X_{i}^{1}X_{i}^{2} -
X_{i}^{1}X_{i}^{2}r_{+} +
X_{i}^{1}r_{-}X_{i}^{2} + X_{i}^{2}r_{+}X_{i}^{1}, \label{x=x}
\end{equation}

\begin{equation}
\{X_{i}^{1}, X_{j}^{2}\} = -r_{+}X_{i}^{1}X_{j}^{2} -
X_{i}^{1}X_{j}^{2}r_{+} +
X_{i}^{1}r_{+}X_{j}^{2} + X_{j}^{2}r_{+}X_{i}^{1}, \label{x>x}
\end{equation}
for $i<j$ and

\begin{equation}
\{A_{i}^{1}, B_{i}^{2}\} = -r_{+}A_{i}^{1}B_{i}^{2} -
A_{i}^{1}B_{i}^{2}r_{+} +
A_{i}^{1}r_{-}B_{i}^{2} + B_{i}^{2}r_{+}A_{i}^{1}.  \label{ab}
\end{equation}
The Poisson brackets (\ref{x=x}--\ref{ab}) admit the restriction to
the set of
functions invariant with respect to simultaneous conjugations of
monodromies:

\begin{equation}
X_{i}^{g} = g X_{i} g^{-1}.  \label{gXg}
\end{equation}
Being restricted they coincide with the canonical Poisson brackets on
the moduli space of flat connections \cite{FoRo}.

{\em Remark 2.} The transformations (\ref{gXg}) may be implemented in
the
quantum algebra ${\cal F}$ as well. To this end one should impose the
quantum
group exchange relation on the transformation parameter $g$:

\begin{equation}
R_{+}g^{1}g^{2}=g^{2}g^{1}R_{+}.  \label{Rgg}
\end{equation}

\begin{definition}
The algebra ${\cal I}_{g,n}$ is an invariant subalgebra of ${\cal
F}_{g,n}$ with respect to
the quantum group action (\ref{gXg}).
\end{definition}

\begin{definition}
The moduli algebra ${\cal M}_{g,n}$ is a factor algebra of ${\cal
I}_{g,n}$ over
the ideal generated by the relation (\ref{ABC}).
\end{definition}

The algebra ${\cal M}$ provides a natural quantization of the moduli
space of flat connections
with marked points. For more extended description one can look the
reference \cite{AGS}.

{\em Remark 3.} One can assign to each link a subalgebra of ${\cal
F}_{g,n}$ generated by
the matrix elements of $X_{i}$.  It is worth mentioning that all
these
subalgebras are
isomorphic to $U_{q}({\cal G})$ \cite{FRT}. For the case of ${\cal
G}=sl(2)$ the isomorphism
may be easily written:

\[ X= \left( \begin{array}{cc}
             K^{2} + q^{-1}(q-q^{-1})^{2}FE & (q-q^{-1})FK^{-1} \\
             (q-q^{-1})K^{-1}E & K^{-2}
\end{array} \right) \]
Here $K,E,F$ are standard generators of $sl_{q}(2)$:

\begin{eqnarray}
KE=qEK, \nonumber \\
KF=q^{-1}FK, \nonumber \\
EF - FE=\frac{K^{2}-K^{-2}}{q-q^{-1}}.
\label{av}
\end{eqnarray}
It is convenient to introduce Gauss decomposition for the quantum
matrix $X$:

\begin{equation}
X=X_{+}X_{-}^{-1}.
\label{aw}
\end{equation}
The diagonal parts of $X_{+}$ and $X_{-}$ are inverse to each other.
The matrices $X_{+}$ and $X_{-}$ satisfy nice exchange relations

\begin{eqnarray}
R_{\pm}X_{\pm}^{1}X_{\pm}^{2}=X_{\pm}^{2}X_{\pm}^{1}R_{\pm},
\nonumber
\\
R_{+}X_{+}^{1}X_{-}^{2}=X_{-}^{2}X_{+}^{1}R_{+}.
\label{ax}
\end{eqnarray}
Matrix elements of $X_{\pm}$ generate upper- and lower-triangular
Borel
subalgebras in $U_{q}({\cal G})$. We shall assume that for any matrix
which satisfies
the exchange relation (\ref{X=X}) one can introduce the Gauss
components
$X_{\pm}$.

To the construction of the integrable model which will appear in the
next section
we introduce the following notations:

\begin{equation}
M_{n+2i-1}=A_{i}, M_{n+2i}=B_{i}^{-1}A_{i}^{-1}B_{i}.  \label{Mn}
\end{equation}
The definition (\ref{Mn}) may be motivated by the observation that
now
the
exchange relations for matrices $M_{i}$ look uniformly:

\begin{eqnarray}
R_{-}M_{i}^{1}R_{-}^{-1}M_{i}^{2}=M_{i}^{2}R_{+}M_{i}^{1}R_{+}^{-1},
\nonumber \\
R_{+}M_{i}^{1}R_{+}^{-1}M_{j}^{2}=M_{j}^{2}R_{+}M_{i}^{1}R_{+}^{-1}.
\label{M>M}
\end{eqnarray}
The last relation as usually holds for $i<j$. It is worth mentioning
that the matrix elements
of the monodromies $M_{i}$ generate a subalgebras in ${\cal F}$. In
principle, one can choose
$M_{i}$ for $i>n$ in a different way:

\begin{equation}
M'_{n+2i-1}=A_{i}B_{i}^{-1}A_{i}^{-1}, M'_{n+2i}=B_{i}^{-1}.
\label{Mnn}
\end{equation}
The relations (\ref{M>M}) are still valid but the subalgebra
generated
by $M'_{i}$ is different
from the subalgebra generated by $M_{i}$.

\section{Trasfer matrix}
\setcounter{equation}{0}

The purpose of this section is to define classical and quantum
completely integrable systems on the moduli space of flat
connections.
To this end one should find a set of functionally independent
commuting Hamiltonians $H_{k}$ ($k=1,\dots ,N/2$) so that its
number is equal to a half of the dimension $N$ of the phase space.
In the framework of the Inverse Scattering method one prefers
to start with a generating function which depends on the complex
parameter $\lambda$ (spectral parameter) and produces the
Hamiltonians
$H_{k}$
in the expansion around some point $\lambda_{0}$ (usually
$\lambda_{0}$ may be chosen to be equal to infinity). So our nearest
goal is to introduce the dependence on the spectral parameter into
the definition of monodromies. For simplicity we shall restrict
ourselves
to the case of $SL(N)$ and even more specifically to its
$N$-dimensional
representation and briefly discuss other situations in the next
section.

Let us introduce a new object

\begin{equation}
M_{i}(\lambda)=M_{i}+\lambda I, \label{Ml}
\end{equation}
where $I$ denotes the $n$ by $n$ unit matrix.
Exchange properties of the matrices $M_{i}(\lambda)$ may be described
as follows.

\begin{theorem}
The matrix elements of spectral dependent monodromies
$M_{i}(\lambda)$
satisfy the following set of quadratic relations:
\begin{eqnarray}
R(\lambda ,\mu)M_{i}^{1}(\lambda)R_{-}^{-1}M_{i}^{2}(\mu)=
M_{i}^{2}(\mu)R_{+}M_{i}^{1}(\lambda)\tilde{R}(\lambda ,\mu),
\nonumber \\
R_{+}M_{i}^{1}(\lambda)R_{+}^{-1}M_{j}^{2}(\mu)=
M_{j}^{2}(\mu)R_{+}M_{i}^{1}(\lambda)R_{+}^{-1}.  \label{Ml>Ml}
\end{eqnarray}
 \end{theorem}
As usual, the last equation is valid for $i<j$. It does not get
changed
comparatively to (\ref{M>M}). As about the first relation, the new
structure
constants $R(\lambda ,\mu)$ and $\tilde{R}(\lambda ,\mu)$ appear.
These are matrices
in the tensor square of the fundamental representation:

\begin{eqnarray}
R(\lambda ,\mu)=\lambda R_{+} - \mu R_{-}, \nonumber \\
\tilde{R}(\lambda ,\mu)=\lambda R_{-}^{-1} - \mu R_{+}^{-1}.
\label{Rl}
\end{eqnarray}
Both $R(\lambda ,\mu)$ and $\tilde{R}(\lambda ,\mu)$ are solutions of
the quantum Yang-Baxter
equation with the spectral parameter:

\begin{equation}
R_{12}(\lambda ,\mu) R_{13}(\lambda ,\nu) R_{23}(\mu ,\nu)=
R_{23}( \mu ,\nu) R_{13}(\lambda ,\nu) R_{12}(\lambda ,\mu).
\label{QYB}
\end{equation}
It is convenient that $R(\lambda ,\mu)$ and $\tilde{R}(\lambda ,\mu)$
may be efficiently
inverted:

\begin{eqnarray}
(\lambda R_{+} - \mu R_{-}) (\lambda R_{+}^{-1} - \mu
R_{-}^{-1})=f(\lambda ,\mu)I, \nonumber \\
(\lambda R_{-}^{-1} - \mu R_{+}^{-1}) (\lambda R_{-} - \mu
R_{+})=f(\lambda ,\mu)I, \label{RRf}
\end{eqnarray}
where the scalar function $f(\lambda ,\mu)$ is equal to:

\begin{equation}
f(\lambda ,\mu)={\lambda}^{2}+{\mu}^{2}-\lambda\mu(q^{2}+q^{-2}).
\label{f}
\end{equation}
Formulae (\ref{RRf}) are based on the important property of $SL(N)$
$R$-matrices in
the $N$-dimensional representation:

\begin{equation}
R_{+}R_{-}^{-1}+R_{-}R_{+}^{-1}=(q^{2}+q^{-2})I.  \label{RRI}
\end{equation}
We postpone the proof of the Theorem 1 until the next section where
the proper technique will be developed and turn to the definition of
the
transfer matrix.

\begin{definition}
The transfer matrix is defined as an
ordered product of spectral dependent momodromies $M_{i}(\lambda)$:

\begin{equation}
T(\lambda)=\prod_{i=1}^{n+2g}{M_{i}(\lambda)}=
M_{1}(\lambda)\dots M_{n+2g}(\lambda).  \label{Tl}
\end{equation}

 \end{definition}

The matrix elements of the transfer matrix satisfy the exchange
relation which is
similar to (\ref{Ml>Ml}):

\begin{equation}
R(\lambda ,\mu)T^{1}(\lambda)R_{-}^{-1}T^{2}(\mu)=
T^{2}(\mu)R_{+}T^{1}(\lambda)\tilde{R}(\lambda ,\mu). \label{Tl>Tl}
\end{equation}
Proof of the formula (\ref{Tl>Tl}) is straightforward. In the course
of commuting
$T(\lambda)$ and $T(\mu)$ one have to use the equality:

\begin{equation}
R_{+}^{-1}R(\lambda ,\mu)=\tilde{R}(\lambda ,\mu)R_{-}.  \label{RRRR}
\end{equation}

The very important consequence of formula (\ref{Tl>Tl}) reads as
follows:

\begin{theorem}
The q-trace of the transfer matrix $T(\lambda)$

\begin{equation}
F(\lambda)=Tr_{q} T(\lambda) \label{Fl}
\end{equation}
provides a family of commuting variables on the moduli space:

\begin{equation}
F(\lambda)F(\mu)=F(\mu)F(\lambda). \label{FF}
\end{equation}

 \end{theorem}
{\em Proof.} Let us multiply the identity (\ref{Tl>Tl}) by
$R_{+}^{-1}$ from the left and
by $(\lambda R_{-} - \mu R_{+})$ from the right. Then we apply the
$q$-trace in the first and in
the second space. The result looks as follows:

\begin{eqnarray}
Tr_{q}^{1,2} R_{+}^{-1}(\lambda R_{+} - \mu
R_{-})T^{1}(\lambda)R_{-}^{-1}T^{2}(\mu)(\lambda R_{-} - \mu R_{+})=
\nonumber \\
=f(\lambda ,\mu) Tr_{q}^{1,2}
R_{+}^{-1}T^{2}(\mu)R_{+}T^{1}(\lambda).
\label{TTa}
\end{eqnarray}
It is easy to check that (\ref{TTa}) may be rewritten as

\begin{equation}
f(\lambda ,\mu) Tr_{q}^{1,2} R_{-}T^{1}(\lambda)R_{-}^{-1}T^{2}(\mu)=
f(\lambda ,\mu) Tr_{q}^{1,2} R_{+}^{-1}T^{2}(\mu)R_{+}T^{1}(\lambda).
\label{TTb}
\end{equation}
Here we have used the fact that $R$-matrices commute with the kernel
of $Tr_{q}^{1,2}$ and
have applied equations (\ref{RRf},\ref{RRRR}). The $q$-trace has a
remarkable property which
enables us to cancel all $R$-matrices both in the l.h.s and in the
r.h.s. of (\ref{TTb}).
If we take into account this fact, formula (\ref{TTb}) implies
commutativity of $F(\lambda)$. Obviously $F(\lambda)$ is invariant
with respect to quantum group
action (\ref{gXg}) and defines a family of commuting Hamiltonians on
the moduli space
of flat connections. This remark completes the proof of {\em Theorem
2} and we turn to
discussion of possible generalizations and applications.

{\em Remark 1}. It is possible to generalize the described procedure
for any simple
Lie group ${\cal G}$ and for any representation $I$. The form of
spectral dependent
monodromies $M_{i}(\lambda)$ will be more sofisticated. It will
consist of $d$ terms,
where $d$ is the number of irreducible representations which appear
in
the tensor
product $I\otimes I^{*}$. It is easy to check that for
$N$-dimensional
representation
of the group $SL(N)$ $d=2$. The matrix $M_{i}(\lambda)$ is always
polynomial in
$\lambda$ but the degree will increase with the number of terms. We
don't give more
details in this paper but we shall use the fact that one can
construct
the quantity
$F^{I}(\lambda)$ for any representation $I$:

\begin{equation}
F^{I}(\lambda)=Tr_{q}M^{I}_{1}(\lambda)\dots M^{I}_{n+2g}(\lambda).
\label{Fi}
\end{equation}
so that all of them commute with each other.

{\em Remark 2}.  Actually, formula (\ref{Fl}) defines a family of
commuting families
parametrized by $n+2g-1$ parameter. In order to construct them, let
us
modify the definition of the transfer matrix:

\begin{equation}
T_{z_{1},\dots
,z_{n+2g}}(\lambda)=\prod_{i=1}^{n+2g}{M_{i}(z_{i}\lambda)}.
\label{Tz}
\end{equation}
The transfer matrix (\ref{Tz}) provides a family of commuting
variables which depends on
the point in $CP^{n+2g-1}$ defined by the parameters $z_{1},\dots
,z_{n+2g}$.
The simultaneous dilatation of the parameters $z_{i}$ is equivalent
to
renormalization
of $\lambda$ and does not change the commuting family. Let us
consider
the simplest example
when the shift parameters $z_{1},\dots ,z_{n+2g}$ are ordered as
follows
$\mid z_{1}\mid\ll\mid z_{2}\mid\ll\dots \ll \mid z_{n+2g}\mid$. Then
the family of commuting
variables corresponding to this choice of shift parameters is
generated by $q$-traces of
matrices $T_{i}$

\begin{equation}
T_{i}=M_{1}M_{2}\dots M_{i}.  \label{Ti}
\end{equation}
This construction resembles the polarization on the moduli space of
$SU(2)$ connections
obtained in \cite{JW}.

{\em Remark 3}. In the classical case we obtain a system of
Hamiltonians in involution.
It means that their pairwise Poisson brackets vanish.  In general,
the
number of independent
functions in our family is less than half of the dimension of the
moduli space . In the case of the group $SU(2)$ the number of
variables is sufficient and the family
of polarizations similar to considered in \cite{JW} may be extracted
from the transfer matrix
(\ref{Tz}).

\section{Equivalence to XXZ magnetic chain}
\setcounter{equation}{0}

Here we shall establish the equivalence between the integrable model
which appeared in the last section and XXZ magnetic chain.
The change of variables on the moduli space
which provides this equivalence is quite important itself because it
furnishes a nice representation for the symplectic structure
\cite{AM}.

Let us introduce new matrix generators $L_{i}(i=1,\dots ,n+2g)$. Each
of them may be
decomposed into the product upper- and lower-triangular multipliers:

\begin{equation}
L_{i}=L_{+}(i)L_{-}(i)^{-1}.  \label{Li}
\end{equation}
We introduce a new algebra.

\begin{definition}
The algebra ${\cal L}_{n+2g}$ is generated by the matrix elements of
$L_{i}$
subject to quadratic relations:

\begin{eqnarray}
R_{-}L_{i}^{1}R_{-}^{-1}L_{i}^{2}=L_{i}^{2}R_{+}L_{i}^{1}R_{+}^{-1},
\nonumber \\
L_{i}^{1}L_{j}^{2}=L_{j}^{2}L_{i}^{1}.  \label{L=L}
\end{eqnarray}
\end{definition}

As we see, the algebra ${\cal L}_{n+2g}$ is isomorphic to
$U_{q}({\cal
G})^{\otimes(n+2g)}$. However, it appears to be very useful. The
following theorem explains the importance of
${\cal L}_{n+2g}$ for the moduli algebra.

\begin{theorem}
The algebra ${\cal L}_{n+2g}$ is isomorphic to the subalgebra in
${\cal F}_{0,n+2g}$ generated by matrix elements of monodromies
$M_{i}$.
\end{theorem}

{\em Proof.} Let us introduce the new set of matrices $K_{i}$ as an
ordered product of
lower-triangular components $L_{-}(i)$:

\begin{equation}
K_{i}=L_{-}(1)\dots L_{-}(i-1), K_{1}=I.  \label{K}
\end{equation}
The isomorphism between ${\cal L}_{n+2g}$ and the corresponding
subalgebra in
${\cal F}_{n+2g}$ is defined by the explicit formulae:

\begin{equation}
M_{i}=K_{i}L_{i}K_{i}^{-1}.  \label{KLK}
\end{equation}
One can easily check that the defining relations (\ref{M>M}) follow
from
(\ref{L=L},\ref{KLK}). The fact that one can always start with a
direct product
of quantum algebras instead of the complicated moduli algebra is in
intimate
relation with the representation of the states in the CS theory as
invariants
in the tensor product of representations of the quantum group.

The important question is what happens to the relation (\ref{ABC})
after the change of
variables. Separating upper- and lower-triangular components one
finds
two relations:

\begin{eqnarray}
L_{+}(1)\dots L_{+}(n+2g)=I, \nonumber \\
L_{-}(1)\dots L_{-}(n+2g)=I.  \label{K+-}
\end{eqnarray}
Now we can clarify the construction of spectral dependent
monodromy:

\begin{equation}
M_{i}(\lambda)=K_{i-1}(L_{+}(i)+\lambda L_{-}(i))K_{i}^{-1}.
\label{GM}
\end{equation}
It appears to be gauge equivalent to the standard $L$-operator for
the
XXZ magnetic chain:

\begin{equation}
L_{i}(\lambda)=L_{+}(i)+\lambda L_{-}(i).  \label{Ll}
\end{equation}
Formulae (\ref{K+-}) assure that the gauge parameter $K_{i}$ is
periodic
on our finite lattice $K_{n+2g+1}=I$. Thus, the trace of the
monodromy
matrix
does not change when we substitute $M_{i}(\lambda)$ instead of
$L_{i}(\lambda)$.
So one may say that the {\em Theorem 2} follows from the
commutativity
theorem
for the trace of monodromy for XXZ magnet. However,
we prefer the formulation of the previous section because the change
of variables from
$M_{i}$ to $L_{i}$ is obviously not canonical.

{\em Theorem 3} enables us to derive exchange relations for
$M_{i}(\lambda)$ in
a simple way.

{\em Proof of Theorem 1}.  It is well known that the $L$-operator of
XXZ model
(\ref{Ll}) satisfies the basic relation:

\begin{equation}
R(\lambda, \mu) L^{1}(\lambda)L^{2}(\mu)=L^{2}(\mu)
L^{1}(\lambda)R(\lambda, \mu).
\label{RLL}
\end{equation}
If we take into account formula (\ref{RLL}) the proof of relations
(\ref{Ml>Ml}) becomes straightforward. In the course of commuting
$M_{i}(\lambda)$
the following identities are especially useful:

\begin{eqnarray}
R_{+}L_{i}^{1}(\lambda)L_{-}^{2}(i)=L_{-}^{2}(i)L_{i}^{1}(\lambda)R_{+
},
\nonumber \\
R_{\pm}K_{i}^{1}K_{i}^{2}=K_{i}^{2}K_{i}^{1}R_{\pm}, \nonumber \\
K_{i}^{1}L_{i}^{2}(\lambda)=L_{i}^{2}(\lambda)K_{i}^{1}.
\end{eqnarray}

\section{Factorization of the moduli algebra}
\setcounter{equation}{0}

The idea of this section is to extend the {\em Theorem 3} to the
algebra ${\cal F}_{g,n}$
corresponding to any genus $g$. This algebra can not be represented
as
a direct product
of several copies of $U_{q}({\cal G})$.  So we have to introduce one
more basic object \cite{AF}.

\begin{definition}
The quantized algebra of functions of the cotangent bundle of the
group $G$
 $Fun(T^{*}G_{q})$ is generated by the matrix elements of matrices
$g$
and $L$
subject to quadratic relations:

\begin{eqnarray}
R_{-}L^{1}R_{-}^{-1}L^{2}=L^{2}R_{+}L^{1}R_{+}^{-1}, \nonumber \\
R_{\pm}g^{1}g^{2}=g^{2}g^{1}R_{\pm}, \nonumber \\
R_{\pm}L_{\pm}^{1}g^{2}=g^{2}L_{\pm}^{1}R_{\pm}.  \label{TG}
\end{eqnarray}
\end{definition}
The algebra $Fun(T^{*}G_{q})$ may be considered as a deformation of
the algebra of differential operators of finite order on the group
$G$. One may
interpret this algebra as an algebra of observables for the quantum
system which
consists of the point particle which moves in the quantum group
$G_{q}$.  From this
point of view it is natural to treat $g$ as coordinates and $L_{\pm}$
as components
of the left momentum. The algebra $Fun(T^{*}G_{q})$ has a unique
representation which
may be realized for example by left multiplication on the algebra
itself. It is convenient
to introduce a matrix $\tilde{L}$:

\begin{equation}
\tilde{L}=g^{-1}L^{-1}g.  \label{N}
\end{equation}
Formula (\ref{N}) introduces the right momentum into the theory:

\begin{eqnarray}
R_{-}\tilde{L}^{1}R_{-}^{-1}\tilde{L}^{2}=\tilde{L}^{2}R_{+}\tilde{L}^
{1}R_{+}^{-1},
\nonumber \\
L^{1}\tilde{L}^{2}=\tilde{L}^{2}L^{1}.  \label{N>N}
\end{eqnarray}
As usual, left and right momenta commute with each other and realize
two independent
copies of $U_{q}({\cal G})$.  If we regard the only representation of
$Fun(T^{*}G_{q})$ as a
representation of the subalgebra generated by the matrix elements of
left and right momenta $L$ and $\tilde{L}$, we discover the regular
representation of $U_{q}(G)$:

\begin{equation}
\Re_{G}=\oplus_{I\in\Im}I\otimes I^{*}, \label{Reg}
\end{equation}
where $\Im$ is the set of irreducible representations of ${\cal G}$.

The algebra ${\cal F}_{g,n}$ may be handled in a way similar to how
we
handle the
algebra ${\cal F}_{0,n+2g}$.

\begin{theorem}
The algebra ${\cal F}_{g,n}$ is isomorphic to the algebra ${\cal
L}_{g,n}=U_{q}({\cal G})
^{\otimes n}\otimes Fun(T^{*}G_{q})^{\otimes g}$.
\end{theorem}

{\em Proof}. We describe the isomorphism explicitly. The algebra
$L_{g,n}$ is generate
by the matrix elements of $L_{1},\dots ,L_{n}$ corresponding to $n$
copies of $U_{q}({\cal G})$
and $(L_{n+1}, g_{1}), (L_{n+3}, g_{2}),\dots ,(L_{n+2g-1},g_{g})$
corresponding to $g$
copies of the algebra $Fun(T^{*}G_{q})$. We have enumerated left
momenta by the indices
$n+2i-1$ and reserved the indices $n+2i$ for right momenta:

\begin{equation}
L_{n+2i}=g_{i}^{-1}L_{n+2i-1}^{-1}g^{i}.  \label{NL}
\end{equation}
As in the previous section we define the variables $K_{i}$:

\begin{equation}
K_{i}=L_{-}(1)\dots L_{-}(i-1), K_{1}=I.  \label{K2}
\end{equation}
The isomorphism which we need looks as:

\begin{eqnarray}
M_{i}=K_{i}L_{i}K_{i}^{-1}, \nonumber \\
A_{i}=K_{n+2i-1}L_{n+2i-1}K_{n+2i-1}^{-1}, \nonumber \\
B_{i}=K_{n+2i-1}g_{i}K_{n+2i}^{-1}.  \label{MAB}
\end{eqnarray}
As usual, the checking of relations (\ref{MAB}) is straightforward.

Let us state an
important property of the algebra ${\cal M}_{g,n}$. The elements

\begin{equation}
c_{i}^{I}=Tr_{q}M^{I}_{i} \label{cM}
\end{equation}
for $i=1,\dots ,n$ belong to the centre of the algebra ${\cal
M}_{g,n}$. Indeed,
due to the first formula (\ref{MAB}) we have

\begin{equation}
c_{i}^{I}=Tr_{q}L^{I}_{i}.  \label{cL}
\end{equation}
It is known that the elements of the type $c_{i}^{I}$ for given $i$
generate the centre of
$U_{q}({\cal G})$. As the algebra ${\cal L}_{g,n}$ includes a direct
product of several copies of
$U_{q}({\cal G})$, we conclude that the elements (\ref{cM}) lie in
the
centre of ${\cal M}_{g,n}$.
Let us remark that the algebra $Fun(T^{*}G_{q})$ has no centre.

Now we are equipped to construct the representation theory of ${\cal
M}_{g,n}$. It is convenient to start with some irreducible
representation of ${\cal F}_{g,n}$, or equivalently ${\cal L}_{g,n}$.
The latter is a direct product and its representations are enumerated
by tuples $(I_{1},\dots ,I_{n})$ of irreducible representations of
${\cal G}$ and may be realized in the space

\begin{equation}
V_{I_{1},\dots ,I_{n}}=I_{1}\otimes \dots \otimes I_{n}\otimes
\Re^{\otimes g}, \label{V}
\end{equation}
where $n$ copies of $U_{q}({\cal G})$ act in the multipliers
$I_{1},\dots ,I_{n}$ and
each copy of $Fun(T^{*}G_{q})$ is realized in $\Re$.

The algebra $U_{q}({\cal G})$ may
be naturally embedded into ${\cal L}_{g,n}$ in the following way:

\begin{equation}
L_{\pm}=\prod_{i=1}^{n+2g}{L_{\pm}(i)}.  \label{GU}
\end{equation}
The algebra of invariants ${\cal I}_{g,n}$ may be defined as a
commutant of the image of
$U_{q}({\cal G})$ embedded by the formula (\ref{GU}). Obviously, the
representation
$V_{I_{1},\dots ,I_{n}}$ gets reducible when we restrict it to the
algebra ${\cal I}_{g,n}$.
Moreover, the representation $V_{I_{1},\dots ,I_{n}}$ may be
decomposed into the direct
sum of irreducible representations of $U_{q}({\cal G})$ and ${\cal
I}_{g,n}$:

\begin{equation}
V_{I_{1},\dots ,I_{n}}=\oplus_{I\in\Im}I\otimes W^{I}_{I_{1},\dots
,I_{n}}.  \label{VIW}
\end{equation}
The algebra ${\cal M}_{g,n}$ is defined as a factoralgebra of ${\cal
I}_{g,n}$ over the ideal defined by relations

\begin{equation}
L_{\pm}=\prod_{i=1}^{n+2g}{L_{\pm}(i)}=I.  \label{GU1}
\end{equation}
So only those representations of the algebra ${\cal I}_{g,n}$ may be
reinterpreted
as representations of ${\cal M}_{g,n}$ where the ideal generated by

\begin{equation}
c^{I}=Tr_{q} L^{I}_{+}(L^{I}_{-})^{-1} \label{cU}
\end{equation}
acts by the trivial representation. There is only one representation
which satisfies this condition in each sum (\ref{VIW}). Namely, one
should pick up a partner of the trivial representation
$W^{e}_{I_{1},\dots ,I_{n}}$ which is actually isomorphic to the
space
of $q$-invariants in the space $V_{I_{1},\dots ,I_{n}}$ regarded as a
representation of $U_{q}({\cal G})$ realized by the formula
(\ref{GU}).  Thus we conclude.

\begin{theorem}
The irreducible representations of the algebra ${\cal M}_{g,n}$ are
enumerated by
tuples $(I_{1},\dots ,I_{n})$ of representations of ${\cal G}$. Each
representation
of this type may be realized in the space

\begin{equation}
W^{e}_{I_{1},\dots ,I_{n}}=Inv_{q}(I_{1}\otimes \dots \otimes
I_{n}\otimes \Re^{\otimes g})
\label{We}
\end{equation}
by the standard formulae which define the action of $U_{q}({\cal G})$
in its representations
$I_{1},\dots ,I_{n}$ and $Fun(T^{*}G_{q})$ in the representation
$\Re$.
\end{theorem}

Let us emphasize that we have proved this {\em Theorem} only for
generic values of the deformation parameter $q$. The most interesting
case of $q$ being a root of unity needs some further consideration
(see also \cite{AGS}).

\section{Discussion}
This Section is devoted to open problems related to quantization of
the Hamiltonian
Chern-Simons theory and to the integrable model on the moduli space
of
flat connections
introduced in Section 3.

The first problem that I would like to mention is the construction of
the representation theory of the moduli algebra for $q$ being a root
of unity.  The difference in comparison with generic $q$ is that we
have to work with quasi-associative algebras instead of associative
if
we want to have nondegenerate scalar product in the representation
space. The other possibility is to deal with the same moduli algebra
as in Section 2 but now it has a nontrivial ideal which should be
factorized out.  The first step in this direction is presented in
\cite{AGS}.

Another problem is related to the question of completeness of the
family of commuting Hamiltonians. For my knowledge, it is possible to
complete this simplest family corresponding to infinite values of
spectral shifts (see Section 3) in order to get polarization.
However,
it would be
interesting to fulfil the same program for arbitrary spectral shifts.
Such construction would provide a new family of polarizations on the
moduli space of flat connections. If we compare this
hypothetical family with polarizations induced by complex structures
on the Riemann surface, we discover that in these two families the
number of parameters corresponding to a marked point coincide (it is
equal to 1), whereas the number of parameters corresponding to a
handle is different (3 in the case of complex structures and only 2
in
our case). This is a hint that it
may be possible to construct $L$-operator of more general form than
the one which we used
in Section 3.  Actually, there we treat a handle as a couple of
marked
points. On the language of complex structures it would mean that we
consider only hyperelliptic surfaces. So, it is
quite possible that some extra spectral shifts may be introduced into
the definition of the transfer matrix.

If we assume that there exists a family of polarizations parametrized
by spectral shifts (at least for $SL(2)$ it is indeed the case) we
face the problem how to deal with different quantizations provided by
these polarizations simultaneously. The experience obtained in the
course of investigation of another family of polarizations on the
same
moduli space \cite{WL} tells that one of possible ways is to
construct
the flat connection on the space of parameters which is designed to
identify canonically the Hilbert spaces obtained upon quantization
starting from different polarizations. It is the
Knizhnik-Zamolodchikov (KZ) equation which provides such a connection
in the case of the family of polarizations related to complex
structures. Now the
exiting question arises what kind of equation one can get instead of
the KZ for the family of polarizations parametrized by spectral
shifts? I hope to return to this question in some other paper.

\section*{Acknowledgments}
I would like to thank L.D.Faddeev, V.V.Fock, H.Grosse, A.Z.Malkin,
A.G.Bytsko and
V.Schomerus for stimulating discussions.

\end{document}